\documentclass{article}   
\usepackage{ijcai07}   
\usepackage{times}   
\usepackage{latexsym}   
\input{psfig}   
   
\newtheorem{THEOREM}{Theorem}[section]
\newenvironment{theorem}{\begin{THEOREM} \hspace{-.85em} {\bf :} }%
                        {\end{THEOREM}}
\newtheorem{LEMMA}[THEOREM]{Lemma}
\newenvironment{lemma}{\begin{LEMMA} \hspace{-.85em} {\bf :} }%
                      {\end{LEMMA}}
\newtheorem{COROLLARY}[THEOREM]{Corollary}
\newenvironment{corollary}{\begin{COROLLARY} \hspace{-.85em} {\bf :} }%
                          {\end{COROLLARY}}
\newtheorem{PROPOSITION}[THEOREM]{Proposition}
\newenvironment{proposition}{\begin{PROPOSITION} \hspace{-.85em} {\bf :} }%
                            {\end{PROPOSITION}}
\newtheorem{DEFINITION}[THEOREM]{Definition}
\newenvironment{definition}{\begin{DEFINITION} \hspace{-.85em} {\bf :} \rm}%
                            {\end{DEFINITION}}
\newtheorem{CLAIM}[THEOREM]{Claim}
\newenvironment{claim}{\begin{CLAIM} \hspace{-.85em} {\bf :} \rm}%
                            {\end{CLAIM}}
\newtheorem{EXAMPLE}[THEOREM]{Example}
\newenvironment{example}{\begin{EXAMPLE} \hspace{-.85em} {\bf :} \rm}%
                            {\end{EXAMPLE}}
\newtheorem{REMARK}[THEOREM]{Remark}
\newenvironment{remark}{\begin{REMARK} \hspace{-.85em} {\bf :} \rm}%
                            {\end{REMARK}}

\newcommand{\thm}{\begin{theorem}}
\newcommand{\lem}{\begin{lemma}}
\newcommand{\pro}{\begin{proposition}}
\newcommand{\dfn}{\begin{definition}}
\newcommand{\rem}{\begin{remark}}
\newcommand{\xam}{\begin{example}}
\newcommand{\cor}{\begin{corollary}}
\newcommand{\prf}{\noindent{\bf Proof:} }
\newcommand{\ethm}{\end{theorem}}
\newcommand{\elem}{\end{lemma}}
\newcommand{\epro}{\end{proposition}}
\newcommand{\edfn}{\bbox\end{definition}}
\newcommand{\erem}{\bbox\end{remark}}
\newcommand{\exam}{\bbox\end{example}}
\newcommand{\ecor}{\end{corollary}}
\newcommand{\eprf}{\bbox\vspace{0.1in}}
\newcommand{\beqn}{\begin{equation}}
\newcommand{\eeqn}{\end{equation}}

\newcommand{\bbox}{\vrule height7pt width4pt depth1pt}

\newcommand{\clm}{\begin{claim}}
\newcommand{\eclm}{\end{claim}}

\newcommand{\rimp}{\Rightarrow}

\newcommand{\band}{\bigwedge}

\newcommand{\inter}{\cap}

\newcommand{\IR}{\mbox{$I\!\!R$}}

\renewcommand{\phi}{\varphi}

\newcommand{\A}{{\cal A}}

\newcommand{\G}{{\cal G}}

\newcommand{\K}{{\cal K}}

\newcommand{\R}{{\cal R}}

\newcommand{\ol}{\setlength{\itemsep}{0pt}\begin{enumerate}}
\newcommand{\eol}{\end{enumerate}\setlength{\itemsep}{-\parsep}}
\newcommand{\ul}{\setlength{\itemsep}{0pt}\begin{itemize}}
\newcommand{\dl}{\setlength{\itemsep}{0pt}\begin{description}}
\newcommand{\edl}{\end{description}\setlength{\itemsep}{-\parsep}}
\newcommand{\eul}{\end{itemize}\setlength{\itemsep}{-\parsep}}

\newcommand{\commentout}[1]{}

\newcommand{\bi}{\begin{itemize}}
\newcommand{\ei}{\end{itemize}}
\newcommand{\be}{\begin{enumerate}}
\newcommand{\ee}{\end{enumerate}}

\newcommand{\Gz}{\G_0}

\renewcommand{\S}{{\cal S}}   
\newcommand{\sfa}{{\sf a}}   
\newcommand{\RCond}{\succeq}   
\newcommand{\doact}{{\sf do}}   
\newcommand{\EQNF}{\mathbf{EQNF}}   
\newcommand{\EQEF}{\mathbf{EQEF}}   
\newcommand{\vecmu}{\vec{\mu}}   
\newcommand{\vecnu}{\vec{\nu}}   
\newcommand{\PS}{\mathcal{PS}}   
\newcommand{\Pgkb}{{\sf Pg}}   
\newcommand{\Pnf}{P^{\mathit{nf}}}   
\newcommand{\vPnf}{\vec{P}^{\mathit{nf}}}   
\newcommand{\vPef}{\vec{P}^{\mathit{ef}}}   
   
\newcommand{\STRAT}{\mathrm{STRAT}}   
    
\newcommand{\EU}{\mathrm{EU}}   
\newcommand{\intension}[1]{[\![ #1 ]\!]_{\PS}}   
\newcommand{\Rrep}{{\bf R}}   
\newcommand{\Skip}{{\sf skip}}   
\newcommand{\stand}[1]{\mbox{\em st}\left (#1 \right )}   
\newcommand{\PM}{\mathit{PM}}   
\newcommand{\ijcai}[1]{\commentout{#1}}   
\newcommand{\ijcaionly}[1]{#1}   
\ijcaionly{\let\citeyear=\shortcite}   
\commentout{   
\newcommand{\band}{\bigwedge}   
\newcommand{\rimp}{\Rightarrow}

\newcommand{\R}{{\cal R}}   
\newcommand{\G}{{\cal G}}   
\newcommand{\A}{{\cal A}}   
\newcommand{\K}{{\cal K}}   
\newcommand{\Gz}{\G_0}   
\renewcommand{\phi}{\varphi}   
\newtheorem{theorem}{Theorem}[section]   
\newtheorem{corollary}{Corollary}[section]   
\newtheorem{lemma}{Lemma}[section]   
\newtheorem{proposition}{Proposition}[section]   
\newtheorem{definition}{Definition}[section]   
\newtheorem{non-theorem}{Non-theorem}[section]   
\newtheorem{claim}{Claim}[section]   
\newcommand{\thm}{\begin{theorem}}   
\newcommand{\lem}{\begin{lemma}}   
\newcommand{\pro}{\begin{proposition}}   
\newcommand{\dfn}{\begin{definition} \rm}   
\newcommand{\rem}{\begin{remark}}   
\newcommand{\xam}{\begin{example}}   
\newcommand{\cor}{\begin{corollary}}   
\newcommand{\prf}{\begin{proof}}   
\newcommand{\ethm}{\end{theorem}}   
\newcommand{\elem}{\end{lemma}}   
\newcommand{\epro}{\end{proposition}}   
\newcommand{\edfn}{\bbox\end{definition}}   
\newcommand{\erem}{\bbox\end{remark}}   
\newcommand{\exam}{\bbox\end{example}}   
\newcommand{\ecor}{\end{corollary}}   
\newcommand{\eprf}{\end{proof}}   
}

\begin{document}   

\title{Characterizing Solution Concepts 
in Games
Using Knowledge-Based   
Programs\\ \mbox{\ \ \ }\\ \mbox{\ \ \ }   
}   

\author{Joseph Y. Halpern%
\thanks{Supported in part by NSF under grants   
CTC-0208535 and ITR-0325453, by ONR under grant N00014-02-1-0455,   
by the DoD Multidisciplinary University Research   
Initiative (MURI) program administered by the ONR under   
grants N00014-01-1-0795 and N00014-04-1-0725, and by AFOSR under grant   
F49620-02-1-0101.}   
\\ Computer Science Department \\ Cornell University,   
U.S.A. \\  e-mail: halpern@cs.cornell.edu    
\And    
Yoram Moses\\   
Department of Electrical Engineering\\   
Technion---Israel Institute of Technology\\   
32000 Haifa, Israel\\   
email: moses@ee.technion.ac.il}   
\maketitle   
   
\vspace{-2in}   
\begin{abstract}   
We show how solution concepts in games    
such as Nash equilibrium, correlated equilibrium, rationalizability, and   
sequential equilibrium   
can be given a   
uniform definition in terms of \emph{knowledge-based programs}.   
Intuitively, all solution concepts are implementations of two   
knowledge-based programs, one appropriate for games represented in   
normal form, the other for games represented in extensive   
form.     
These knowledge-based programs can be viewed as embodying rationality.   
The representation works even if (a) information sets do not   
capture an agent's knowledge, (b) uncertainty is not    
represented by probability,    
or   
(c) the underlying game is not common knowledge.     
\end{abstract}   
   
\section{Introduction}   
Game theorists represent games in two standard ways: in \emph{normal form},   
where    
each agent simply    
chooses a strategy, and in \emph{extensive form},   
using game trees, where    
the agents make    
choices over time.  An extensive-form representation has the   
advantage that it describes the    
dynamic    
structure of the   
game---it  explicitly represents the sequence of decision problems   
encountered by agents.  However, the extensive-form representation    
purports    
to do more than just describe the structure of the game; it also    
attempts   
to represent the information    
that players have    
in the game, by the use of   
\emph{information sets}.  Intuitively, an information set consists of a   
set of nodes in the game tree where a player has the same information.     
However,    
as Halpern \citeyear{Hal15} has pointed out,    
information sets may not adequately represent a player's information.   
   
Halpern makes this point by considering the following single-agent game   
of imperfect recall, originally    
presented   
by Piccione and Rubinstein   
\citeyear{PR97}:    
\begin{figure}[htb]   
\centerline{\psfig{figure=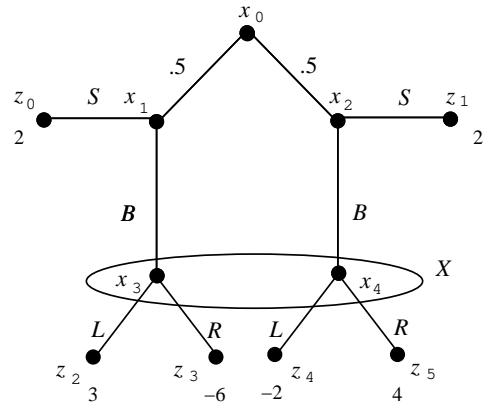,height=2.2in}}   
\caption{A game of imperfect recall.}   
\label{fig2}   
\end{figure}   
The game starts with nature moving either left or right, each with    
probability $1/2$.  The    
agent    
can then either stop the game   
(playing move $S$) and get a payoff of 2, or continue, by playing move   
$B$.  If he continues, he gets a high    
payoff if he matches nature's move, and a low payoff otherwise.   
Although he originally knows nature's move, the information set that   
includes the nodes labeled $x_3$ and $x_4$ is intended to indicate that   
the player forgets whether nature moved left or right after moving $B$.   
Intuitively, when he is at the information    
set~$X$,    
the agent is not   
supposed to know whether he is at $x_3$ or at $x_4$.   
   
It is not hard to show that the strategy that maximizes expected utility   
chooses action $S$ at node $x_1$, action $B$ at node   
$x_2$, and action $R$ at the information set $X$ consisting of $x_3$ and   
$x_4$.  Call this strategy $f$.  Let $f'$ be the strategy of choosing   
action $B$ at $x_1$, action $S$ at $x_2$, and $L$ at $X$.   
Piccione and Rubinstein argue that if node $x_1$ is reached, the   
player   
should reconsider, and decide to switch from $f$ to $f'$.   
As Halpern points out, this is indeed true, provided that the    
player   
knows   
at each stage of the game   
what strategy he is currently using.    
However, in that case, if the    
player   
is using $f$ at the information set,   
then he knows that he is at node $x_4$; if he has switched and is   
using $f'$, then he knows that he is at $x_3$.  So, in this setting, it   
is no longer the case that the    
player   
does not know whether he is at   
$x_3$ or $x_4$ in the information set; he can infer which    
state    
he is at from the strategy he is using.   

In game theory, a \emph{strategy} is taken to be a function from   
information sets to actions.     
The intuition behind this is   
that, since an agent cannot tell the nodes in an information set apart,   
he must do the same thing at    
all these nodes.  But this example shows that if the agent has imperfect   
recall but can switch strategies, then he can arrange to do different   
things at different nodes in the same information set.   
As Halpern \citeyear{Hal15} observes,    
`\,``situations that [an agent] cannot   
distinguish'' and ``nodes in the same information set'' may be two quite   
different notions.'  He suggests using the game tree to describe   
the structure of the game, and using the runs and systems framework   
\ijcai{   
introduced by    
Moses \citeyear{Mosthesis} and by    
Halpern and Fagin \citeyear{HFfull}, and further developed by   
Fagin et al.~\citeyear{FHMV}}   
\ijcaionly{\cite{FHMV}}   
to describe the agent's information.     
The idea is that an agent has an internal \emph{local state} that   
describes all the information that he has.  A strategy (or   
\emph{protocol} in the language of \cite{FHMV}) is a function from local   
states to actions.  Protocols capture the intuition that what an agent   
does can depend only what he knows.  But now an agent's knowledge is   
represented by its local state, not    
by    
an information set.  Different   
assumptions about what agents know (for example,    
whether    
they know their current strategies) are   
captured by running the same protocol in different \emph{contexts}.   
If the information sets appropriately represent an agent's knowledge in   
a game, then we can identify local states    
with    
information sets.  But, as   
the example above shows, we cannot do this in general.   

A number of \emph{solution concepts} have been considered   
in the game-theory literature, ranging from Nash equilibrium
and \emph{correlated equilibrium} to refinements of Nash equilibrium
such as \emph{sequential    
equilibrium} and weaker notions such as \emph{rationalizability} (see
\cite{OR94} for an overview).     
The fact that game trees represent both the game and the players'   
information has proved critical in defining solution concepts in   
extensive-form games.  Can we still represent solution concepts in a   
useful way using  runs and systems to represent a player's information?   
As we show here,  not only can we do this, but we can do it in a way   
that gives deeper insight into solution concepts.   
Indeed, all the standard solution concepts    
in the literature   
can be understood as instances of a single     
\emph{knowledge-based (kb) program} \cite{FHMV,FHMV94}, which captures   
the underlying intuition that a player should make a best
response, given her beliefs.    
The differences between solution concepts arise from 
running the kb program in different contexts.   
   
In a kb program,    
a player's actions depend explicitly on the player's knowledge.   
For example, a kb program could have a test that   
says ``If you don't know that Ann received the information, then   
send her a message'',     
which can be written 
$$\quad \mathbf{if}\ \neg B_i(\mathrm{Ann\  received\ info}) \    
\mathbf{then}\  \mathrm{send\ Ann\ a\  message}.$$   
This kb program has the form of a standard {\bf{if}} \ldots {\bf{then}}   
statement, except that the test in the {\bf{if}} clause is a test on $i$'s   
knowledge (expressed using the modal operator $B_i$ for belief; see   
Section~\ref{sec:background} for a discussion of the use of knowledge   
vs.~belief).   
   
Using such tests for knowledge    
allows   
us to   
abstract away from low-level details of how the knowledge is obtained.   
Kb programs have been applied to a number of problems in the computer   
science literature (see \cite{FHMV} and the references therein).  To see   
how they can be applied to understand equilibrium,    
given a game $\Gamma$ in normal form, let   
$\S_i(\Gamma)$ consist of all the    
pure   
strategies for player   
$i$ in $\Gamma$.     
Roughly speaking, we want a kb  program that says that if player $i$   
believes that she is about to perform strategy $S$ (which we express   
with the formula $\doact_i(S$)), and she believes that she would not do   
any better with another strategy, then she should indeed go ahead and run   
$S$.      
This test can be viewed as embodying rationality.   
There is a subtlety in expressing the statement ``she would not   
do any better with another strategy''.  We express this by saying ``if   
her expected utility, given that she will use strategy $S$, is $x$, then    
her expected utility if she were to use strategy $S'$ is at most $x$.''   
The ``if she were to use $S'$'' is a \emph{counterfactual} statement.   
She is planning to use strategy $S$, but is contemplating what would   
happen if she were to do something counter to fact, namely, to use $S'$.   
Counterfactuals have been the subject    
of intense study in the philosophy literature (see, for example,   
\cite{Lewis73,Stalnaker68}) and, more recently, in the game theory   
literature (see, for example, \cite{Aumann95,Hal28,Samet96}).   
We write the counterfactual ``If $A$ were the case then $B$ would be   
true'' as ``$A \RCond B$''.  Although this statement involves an ``if   
\ldots then'', the semantics of the counterfactual implication $A \RCond   
B$ is quite different from the material implication $A \rimp B$.  In   
particular, while $A \rimp B$ is true if $A$ is false, $A \RCond B$   
might not be.     
   
With this background, consider the following kb program for player $i$:   
$$\begin{array}{ll}   
\mathbf{for\ each}\  \mathrm{strategy}~S \in \S_i(\Gamma) \ \mathbf{do}\\   
\quad \mathbf{if}\ B_i(\doact_i(S) \land \forall x(\EU_i = x \rimp\\   
\qquad \band_{S' \in \S_i(\Gamma)}   
(\doact_i(S') \RCond (\EU_i \le x))))\  \mathbf{then}\  S.   
\end{array}$$   
This kb program is meant to capture the intuition above.     
Intuitively, it says that if player $i$ believes that she is   
about to perform strategy $S$ and, if her    
expected utility is $x$, then    
if she were to   
perform another strategy $S'$,    
then    
her expected utility would be    
no   
greater than $x$,    
then she   
should perform strategy $S$.     
Call this kb program $\EQNF^\Gamma$ (with the individual instance for    
player~$i$ denoted by $\EQNF^\Gamma_i$).   
As we show, if all players    
follow $\EQNF^\Gamma$,    
then they end up   
playing some type of equilibrium.  Which type of equilibrium they play   
depends on the context.   
Due to space considerations, we focus on three examples in this   
abstract.    
If the players have a   
common prior on the joint strategies being used, and this common prior   
is such that players' beliefs are independent of the    
strategies   
they use,   
then they play a Nash equilibrium.     
Without this independence assumption, we get a \emph{correlated equilibrium}.   
On the other hand, if players have possibly different priors on the   
space of strategies, then this kb program defines \emph{rationalizable}   
strategies \cite{Ber84,Pearce84}.   

To deal with extensive-form games, we need a slightly different kb   
program, since agents choose moves, not strategies.  Let   
$\EQEF^\Gamma_i$ be the following program, where    
$\sfa \in \PM$ denotes that $\sfa$ is a move that is currently possible.   
$$\begin{array}{ll}   
\mathbf{for\ each}\ \mathrm{move}\ \sfa \in \PM\ \mathbf{do}\\   
\quad \mathbf{if}\ B_i(   
\doact_i(\sfa) \land \forall x ((   
\EU_i = x) \rimp \\   
\quad \quad \band_{\sfa' \in \PM}    
(\doact_i(\sfa') \RCond (\EU_i \le x)))) \  \mathbf{then}\  \sfa.   
\end{array}   
$$   
Just as $\EQNF^\Gamma$ characterizes equilibria of a game   
$\Gamma$ represented in normal form, $\EQEF^\Gamma$ characterizes   
equilibria of a game represented in extensive form.   
We give one example here: sequential equilibrium.     
To capture sequential equilibrium, we need to    
assume that information sets do correctly describe an agent's   
knowledge.     
If we drop this assumption, however, we can distinguish    
between the two equilibria for the game described in   
Figure~\ref{fig2}.     

All these solution concepts are based on expected utility.   
But we can also consider solution concepts based on other decision   
rules.  For example, Boutilier and Hyafil \citeyear{BH04} consider   
\emph{minimax-regret} equilibria, where each player uses a strategy that   
is a best-response in a minimax-regret sense to the choices of the other   
players.  Similarly, we can use    
\emph{maximin equilibria}   
\cite{AB06}.  As pointed out by Chu and Halpern   
\citeyear{CH03a}, all these decision rules can be viewed as instances of   
a generalized notion of expected utility, where uncertainty is   
represented by a \emph{plausibility measure}, a generalization of a   
probability measure, utilities are elements of an arbitrary partially   
ordered space, and plausibilities and utilities are combined using   
$\oplus$ and $\otimes$, generalizations of $+$ and $\times$.     
We show    
in the full paper    
that, just by interpreting ``$\EU_i=u$'' appropriately, we can capture   
these    
more exotic   
solution concepts as well.   
Moreover, we can capture solution concepts in games where the game   
itself is not    
common knowledge,   
or where agents are not aware of all   
moves available, as discussed by Halpern and R\^ego \citeyear{HR06}.   

Our approach thus provides a   
powerful tool for representing   
solution concepts, which works even if (a) information sets do not   
capture an agent's knowledge, (b) uncertainty is not    
represented by probability,    
or    
(c) the underlying game is not common knowledge.     

The rest of this paper is organized as follows.  In   
Section~\ref{sec:background}, we review the relevant background on game   
theory and knowledge-based programs.  In Section~\ref{sec:mainresults},   
we show that $\EQNF^{\Gamma}$ and $\EQEF^{\Gamma}$ characterize Nash   
equilibrium, correlated equilibrium, rationalizability, and sequential   
equilibrium in a game $\Gamma$ in the appropriate contexts.  We conclude   
in Section~\ref{sec:conclusions} with a discussion of how our results   
compare to other characterizations of solution concepts.   

\section{Background}\label{sec:background}   
   
In this section, we review the relevant background    
on games and   
knowledge-based programs.  We describe only    
what we need for proving    
our results.  The reader is encouraged to consult    
\cite{OR94} for more on game theory,    
\cite{FHMV,FHMV94} for more on knowledge-based programs without   
counterfactuals, and \cite{HM98a} for more on adding counterfactuals to   
knowledge-based programs.   

\commentout{   
\subsection{Lexicographic Probability Systems}   
An LPS  is just a pair  $(W,\vecmu)$, where $W$ is a sample space and   
$\vecmu = (\mu^0, \ldots, \mu^k)$ is a sequence of probability measures   
on $W$.  In    
our setting, $W$ is finite and the   
probability measures $\mu^j$ are defined on all subsets of $W$.   
Intuitively, $\mu^0$ is the most significant measure, $\mu^1$ is the   
next most significant measure, and so on.     
We identify a probability measure $\mu$ with the LPS $(\mu)$ of length   
1.   
Given a real-valued random   
variable $f$ on $W$, we define the expected value of $f$ with respect to   
a sequence~%
$\vecmu$, denoted $E_{\vecmu}[f]$, to be the tuple $(E_\mu^0[f]),   
\ldots, E_\mu^k[f])$, where the    
$j^{\rm th}$     
component of the tuple is the   
expected value of $f$ with respect to $\mu^j$.   
   
BBD define conditioning    
with respect to an LPS   
as follows.  Given an LPS $\vecmu =   
(\mu^0, \ldots, \mu^k)$ on a   
space $W$ and $U \subseteq W$   
such that $\mu^j(U) > 0$ for some index $j$, let $\vecmu|U =   
(\mu^{m_0}(\cdot \mid U), \ldots, \mu^{m_h}(\cdot \mid U))$,   
where $(m_0, m_1, \ldots, m_h)$ is the subsequence of all indices    
for which the probability of $U$ is positive.     
Note that   
$\vecmu|U$ is undefined if $\mu_j(U) = 0$ for all $j$.   
}   

\subsection{Games and Strategies}   
A game in \emph{extensive form} is described by a game tree.   
Associated with each non-leaf node or history is either a player---the   
player whose move it is at that node---or nature (which can make a   
randomized move).  The nodes where a player $i$ moves are further   
partitioned into \emph{information sets}.   
With each run or maximal history $h$     
in the game tree and    
player $i$ we can associate $i$'s utility, denoted $u_i(h)$, if that   
run is played. A \emph{strategy} for player $i$ is a (possibly   
randomized) function from $i$'s    
information sets to actions.    
Thus a strategy for player $i$ tells player $i$ what to do at each node   
in the game tree where $i$ is supposed to move.   
Intuitively, at all the nodes that player $i$ cannot tell apart, player   
$i$ must do the same thing.     
A \emph{joint strategy} $\vec{S} = (S_1,   
\ldots, S_n)$ for the players determines a distribution over   
paths in the game tree.  A \emph{normal-form game} can be viewed as a   
special case of an extensive-form game where    
each player makes only one move, and    
all players move   
simultaneously.

\subsection{Protocols, Systems, and Contexts}   
   
To explain kb programs, we must first describe standard protocols.   
We assume that, at any given point in time, a player in a   
game is in some \emph{local state}.     
The local state could include the   
history of the game up to this point, the strategy being used by the   
player, and perhaps some other features of the player's type, such as   
beliefs about the strategies being used by other players.  A global   
state is a tuple consisting of a local state for each player.   
   
A \emph{protocol} for player $i$ is a function from player $i$'s local   
states to actions.  For    
ease of exposition, we consider only deterministic   
protocols, although    
it is relatively straightforward to model    
randomized protocols---corresponding to mixed strategies---as   
functions from local states to distributions over actions.     
Although we restrict to deterministic protocols, 
we deal with mixed strategies by considering distributions over pure
strategies.  

A \emph{run} is a sequence of global states;    
formally, a run is a function from times to global states.   
Thus, $r(m)$ is the global state in run $r$ at time $m$.   
A \emph{point} is a pair $(r,m)$ consisting of a run $r$ and time~$m$.    
Let $r_i(m)$ be $i$'s local state at the point $(r,m)$;   
that is, if    
$r(m) = (s_1, \ldots, s_n)$,    
then $r_i(m) = s_i$.   
A \emph{joint protocol} is    
an assignment of    
a protocol for each player; essentially,   
a joint protocol is a joint strategy.   At each point, a joint   
protocol $\vec{P}$ performs a \emph{joint action}    
$(P_1(r_1(m)), \ldots, P_n(r_n(m)))$,    
which changes the global state.     
Thus, given an initial global state, a joint protocol $\vec{P}$    
generates a    
(unique)    
run, which can be thought of as an execution of $\vec{P}$.   
The runs in a normal-form game involve only one round and two time   
steps: time 0 (the initial state) and time 1,    
after the    
joint strategy has been executed.    
(We assume that the   
payoff is then represented in the player's local state at time 1.)   
In an extensive-form game, a run is again characterized by the   
strategies used, but now the length of the run depends on the path of   
play.     
\ijcai{   
We remark that in typical extensive-form games at most one player    
moves at any given point. This would correspond to at most one player    
having a local state that signals that the player should move at a state,    
and having $P_j(r_j(m))$ be    
the ``{\em null}'' move $\Skip$ for all other players.    
A player performing $\Skip$ doesn't affect the state of the system in any way.   
}   

\ijcai{In order to  study the properties of a joint protocol, and the
knowledge     
that agents have when running it, we consider the set of    
runs of the protocol,   
together with a possibly different    
probability measure    
on runs for   
each player.}
A \emph{probabilistic system} is a tuple    
$\PS = (\R,\vecmu)$, where $\R$ is a set of runs and    
$\vecmu=(\mu_1, \ldots, \mu_n)$ associates a probablity $\mu_i$    
on the runs of $\R$ with each player~$i$.
Intuitively, $\mu_i$ represents player $i$'s prior beliefs.     
In the special case where $\mu_1 = \cdots = \mu_n = \mu$,    
the players have a    
\emph{common prior}   
$\mu$ on $\R$.    
In this case, we    
write just $(\R,\mu)$.   

We are interested in the system corresponding to a   
joint protocol~$\vec{P}$.  To    
determine this system,   
we need to describe the setting   
in which $\vec{P}$ is being executed.  For our purposes, this setting can be   
modeled   
by a set $\G$ of global states, a subset $\Gz$ of $\G$   
that describes the possible    
\emph{initial}    
global states, a set $\A_s$ of   
possible joint actions at each global state $s$, and    
$n$ probability measures on $\Gz$, one for each player.  Thus, a   
\emph{probabilistic context} is a tuple    
$\gamma = (\G, \Gz, \{\A_s: s \in \G\},\vecmu)$.%
\footnote{We are implicitly assuming that the global state that results    
from performing a joint action in $\A_s$ at the global state~$s$ is    
unique and   
obvious; otherwise, such information would also appear in the context,   
as in the general framework of \cite{FHMV}.}   
A joint protocol $\vec{P}$ is \emph{appropriate}   
for such a context~$\gamma$ if,    
for every global state $s$, the joint actions that $\vec{P}$    
can generate are in $\A_s$.    
When $\vec{P}$ is appropriate for~$\gamma$, we abuse notation slightly and    
refer to~$\gamma$ by specifying only the pair $(\Gz,\vecmu)$.    
A protocol $\vec{P}$ and    
a context $\gamma$ for which $\vec{P}$ is appropriate    
generate    
a system; the system depends on the initial states and probability measures   
in $\gamma$. Since these are all that matter, we typically    
simplify the description of a context by omitting    
the set $\G$ of global states and the sets $\A_s$ of global actions.   
Let $\Rrep(\vec{P},\gamma)$ denote the system generated by joint protocol   
$\vec{P}$ in context    
$\gamma$.  If $\gamma = (\Gz, \vecmu)$, then   
$\Rrep(\vec{P},\gamma) = (\R,\vecmu')$, where $\R$   
consists of    
a the run $r_{\vec{s}}$ for each initial state $\vec{s} \in \Gz$,     
where   
$r_{\vec{s}}$ is the run generated by $\vec{P}$    
when    
started in state   
$\vec{s}$,      
and $\mu_i'(r_{\vec{s}}) = \mu_i(\vec{s})$, for $i = 1,   
\ldots, n$.     
   
A    
probabilistic    
system $(\R,\vecmu')$ is \emph{compatible} with a   
context $\gamma = (\Gz,\vecmu)$   
if (a) every initial state in $\Gz$ is the   
initial state of some run in $\R$, (b) every run is the run of some protocol   
appropriate for    
$\gamma$, and (c) if $\R({\vec{s}})$ is the set of runs in   
$\R$ with initial global state $\vec{s}$, then $\mu_j'(\R(\vec{s})) =    
\mu_j(\vec{s})$, for $j = 1, \ldots, n$.     
Clearly $\Rrep(\vec{P},\gamma)$ is compatible with $\gamma$.   

We can think of the context as describing background    
information.  In    
distributed-systems applications, the context also   
typically includes information about message delivery.     
For example,   
it may determine whether all messages sent are received in    
one   
round, or whether they may take up to, say, five rounds.    
Moreover,    
when this is not obvious,    
the context specifies how actions transform the global state;   
for example, it describes what happens if in the same joint action two   
players attempt to    
modify the same memory cell. Since such issues do not   
arise in the games we consider, we ignore these facets of contexts here.   
For simplicity,    
we consider only contexts where   
each initial state corresponds to a particular joint strategy of~$\Gamma$.    
That is, $\Sigma^\Gamma_i$ is a set of local states for   
player $i$ indexed by    
(pure)    
strategies.     
The set $\Sigma^\Gamma_i$ can be viewed as describing   
$i$'s types; the state $s_S$ can the thought of as the initial state   
where player $i$'s type is such that he plays $S$ (although we stress   
that this is only intuition; player $i$ does not    
\emph{have}   
 to play $S$ at the   
state $s_S$).  Let $\Gz^\Gamma =   
\Sigma^\Gamma_1 \times \ldots \times \Sigma^\Gamma_n$.     
We will be interested in contexts where the set of   
initial global states is a    
subset $\Gz$ of $\Gz^\Gamma$.   
In a normal-form game, the only actions   
possible for player $i$ at an initial global state    
amount to choosing a pure strategy,    
so the joint actions are joint strategies; no actions are   
possible at later times.     
For an extensive-form game, the possible moves are described by the game   
tree.    
We say    
that    
a context for an extensive-form game is \emph{standard} if the   
local states have the form $(s,I)$, where $s$ is the initial state   
and $I$ is the   
current information set.  In a standard context, an agent's knowledge is   
indeed described by the information set.  However, we do not require a   
context to be standard.  For example, if an agent is allowed to switch   
strategies, then the local state could include the history of   
strategies used.  In such a context, the agent in the game of   
Figure~\ref{fig2} would know more than just what is in the information   
set, and would want to switch strategies.   

\subsection{Knowledge-Based Programs}   
A \emph{knowledge-based program} is a syntactic object.  For our   
purposes, we can take a knowledge-based program for player $i$ to have   
the form    
$$\begin{array}{l}   
{\bf if }~\kappa_1 ~{\bf then} ~{\sfa_1}\\   
{\bf if }~\kappa_2 ~{\bf then} ~{\sfa_2}\\   
\dots,\\   
\end{array}   
$$   
where    
each   
$\kappa_j$ is a Boolean combination of formulas of the form $B_i \phi$,   
in which    
the $\phi$'s can have nested occurrences of    
$B_\ell$ operators and   
counterfactual implications.    
We assume that the tests $\kappa_1,   
\kappa_2, \ldots$ are mutually exclusive and exhaustive, so that exactly    
one  will evaluate to true in any given instance.    
The program $\EQNF_i^\Gamma$ can be written in this form by simply replacing 
the $\mathbf{for \ \ldots \ do}$ statement by one line for each pure   
strategy in $\S_i(\Gamma)$; similarly for $\EQEF_i^\Gamma$.      

We want to associate a protocol with a kb program.  Unfortunately, we   
cannot ``execute'' a kb program    
as   
we can a protocol.   
How the kb program executes depends on the outcome of    
tests~$\kappa_j$.     
Since the   
tests involve beliefs and counterfactuals, we need to interpret them with   
respect to a system.  The idea is that a kb program    
$\Pgkb_i$ for player~$i$    
and a probabilistic system $\PS$ together determine a protocol $P$   
for player $i$.     
Rather than giving the general definitions   
(which can be found in \cite{HM98a}),   
we just   
show how they work in the two kb programs we consider in this paper:   
$\EQNF$ and $\EQEF$.   

Given a system $\PS = (\R,\vecmu)$, we    
associate with each formula $\phi$ a set $\intension{\phi}$ of points in   
$\PS$.     
Intuitively, $\intension{\phi}$   
is the set of points of~$\PS$   
where the  formula $\phi$ is true. We need a   
little notation:    
\begin{itemize}   
\item If $E$ is a set of points in $\PS$, let $\R(E)$   
denote the set of runs going through points in $E$; that is $\R(E) =   
\{r: \exists m((r,m) \in E)\}$.     
\item Let $\K_i(r,m)$ denote the set of points that $i$ cannot   
distinguish from $(r,m)$:    
 $\K_i(r,m) = \{(r',m'): (r'_i(m') = r_i(m)\}$.   
Roughly speaking, $\K_i(r,m)$ corresponds to $i$'s information set at the   
point $(r,m)$.   
\item Given a point $(r,m)$ and a player $i$, let $\mu_{(i,r,m)}$ be the   
probability measure that results from conditioning $\mu^i$    
on $\K_i(r,m)$, $i$'s information at $(r,m)$.     
We cannot condition on $\K_i(r,m)$ directly: $\mu^i$ is a probability   
measure on    
runs, and $\K_i(r,m)$ is a set of points.  So we actually condition, not   
on $\K_i(r,m)$, but on $\R(\K_i(r,m))$, the set of runs going through   
the points in $\K_i(r,m)$.     
Thus,  $\mu_{i,r,m} = \mu^i \mid \R(\K_i(r,m))$.   
(For the purposes of this abstract, we do not specify $\mu_{i,r,m}$ if   
$\mu_i(\R(\K_i(r,m))) = 0$.     
It turns out not to be relevant to our discussion.)   
\end{itemize}   

The kb programs we consider in this paper use a limited collection of   
formulas.     
We now can  define $\intension{\phi}$ for 
the formulas we consider that do not involve counterfactuals.   
\begin{itemize}   
\item In a system $\PS$ corresponding to a normal-form game $\Gamma$, if   
 $S \in \S_i(\Gamma)$, then    
$\intension{\doact_i(S)}$ is the set of initial points $(r,0)$ such   
that player $i$ uses strategy $S$ in run $r$.     
\item Similarly, if $\PS$ corresponds to an extensive-form game,   
then $\intension{\doact_i(\sfa)}$ is the set of points $(r,m)$ of $\PS$    
at which~$i$ performs action $\sfa$.   
\item Player $i$    
believes a formula $\phi$ at a point $(r,m)$ if    
the event corresponding to formula $\phi$ has    
probability 1 according to    
$\mu_{i,r,m}$.  That is, $(r,m) \in \intension{B_i \phi}$ if   
$\mu_i(\R(\K_i(r,m)) \ne 0$ (so that conditioning on $\K_i(r,m)$ is   
defined) and    
$\mu_{i,r,m}(\intension{\phi} \inter \K_i(r,m)) = 1$.   
\item    
With every run $r$ in the systems we consider, we can   
associate the joint (pure) strategy $\vec{S}$ used in~$r$.%
\footnote{If we allow players to change strategies during a run, then we
will in general have different joint strategies at each point in a run.
For our 
theorems in the next section, we restrict to contexts where players do
not change strategies.}
This pure strategy determines the history in the game, and thus   
determines player $i$'s utility.  Thus, we can associate with every   
point $(r,m)$ player $i$'s \emph{expected} utility at $(r,m)$, where the   
expectation is taken with respect to the probability $\mu_{i,r,m}$.   
If $u$ is a    
real number,   
then $\intension{\EU_i = u}$ is the set of points    
where player   
$i$'s expected utility is $u$;   
$\intension{\EU_i \le u}$ is defined similarly.   
\item    
Assume that $\phi(x)$ has no occurrences of $\forall$.
Then   
$\intension{\forall x \phi(x))} = \inter_{a \in   
\IR}\intension{\phi[x/a]}$, where $\phi[x/a]$ is the result of replacing   
all occurrences of $x$ in $\phi$ by $a$.  That is, $\forall x$ is just   
universal quantification over $x$, where $x$ ranges over the reals.   
This quantification arises for us when $x$ represents a utility, so   
that $\forall x \phi(x)$ is saying that $\phi$ holds for all choices of   
utility.     
\end{itemize}   

We now give the semantics of formulas involving counterfactuals.  Here   
we consider only a restricted class of such formulas, those    
where    
the counterfactual only occurs in the form $\doact_i(S) \RCond \phi$, which   
should be read as ``if $i$ were to use strategy $S$, then $\phi$ would   
be true.     
Intuitively, $\doact_i(S) \RCond \phi$ is true at a   
point $(r,m)$    
if $\phi$ holds in a world that differs from $(r,m)$ only in that $i$ uses the   
strategy $S$.    
That is, $\doact_i(S) \RCond \phi$ is true at $(r,m)$ if $\phi$ is true at   
the point $(r',m)$ where, in run $r'$,   
player $i$ uses strategy $S$ and all the other players use the same   
strategy that they do at $(r,m)$.   
(This can be viewed as an instance of the general semantics for   
counterfactuals used in the philosophy literature   
\cite{Lewis73,Stalnaker68} where $\psi \RCond \phi$ is taken to be true   
at a world $w$ if $\phi$ is true at all the worlds $w'$ closest to $w$   
where $\psi$ is true.)   
Of course, if $i$ actually uses   
strategy $S$ in run $r$, then $r' = r$.     
Similarly, in an extensive-form game $\Gamma$,   
the closest point to   
$(r,m)$ where $\doact_i(\sfa')$ is true (assuming that $\sfa'$ is an   
action that $i$ can perform in the local state $r_i(m)$) is the point   
$(r',m)$ where all players other than player $i$ use the same    
protocol in $r'$ and $r$, and $i$'s protocol in $r'$ agrees with $i$'s   
protocol in $r$ except at the local state $r_i(m)$,   
where  $i$ performs action $\sfa'$.  Thus,   
$r'$ is the run that results    
from player $i$ making a single deviation (to~$\sfa'$ at time~$m$)   
from the protocol she uses in $r$,    
and all other players use the same protocol as in $r$.

There is a problem with this approach.     
There is no guarantee that, in general, such a    
closest point $(r',m)$ exists in the system $\PS$.     
To deal with this problem, we restrict    
attention   
to a class of systems where this   
point is guaranteed to exist.  A system $(\R, \vecmu)$ is \emph{complete   
with respect to context $\gamma$} if    
$\R$   
includes every   
run generated by a protocol appropriate for context $\gamma$.  In   
complete systems, the closest point $(r',m)$ is guaranteed to exist.   
For the remainder of the paper, we evaluate formulas only with respect   
to complete    
systems.   
In a complete system $\PS$, we define $\intension{\doact_i(S) \RCond   
\phi}$ to consist of all the points $(r,m)$ such that the closest point   
$(r',m)$ to $(r,m)$ where $i$ uses strategy $S$ is in   
$\intension{\phi}$.  The definition of $\intension{\doact_i(\sfa) \RCond    
\phi}$    
is similar.    
We say that a complete system $(\R',\vecmu')$   
\emph{extends}    
$(\R,\vecmu)$ if $\mu_j$ and $\mu'_j$ agree on   
$\R$ (so that    
$\mu'_j(A) = \mu_j(A)$) for all $A \subseteq \R$) for $j = 1, \ldots, n$.     
   
Since each formula $\kappa$ that appears as a test in a kb program    
$\Pgkb_i$ for player $i$ is a Boolean combination of formulas of the form   
$B_i\phi$, it is easy to check    
that if    
$(r,m)\in\intension{\kappa}$, then $\K_i(r,m)\subseteq\intension{\kappa}$.     
In other words,    
the truth of $\kappa$    
depends only on $i$'s local state.  Moreover, since the tests are   
mutually exclusive and exhaustive, exactly one of them holds in each   
local state.    
Given a system $\PS$, we take the protocol $\Pgkb_i^{\PS}$ to be such   
that   
$\Pgkb_i^{\PS}(\ell) = \sfa_j$ if, for some point $(r,m)$ in $\PS$ with   
$r_i(m) = \ell$, we have $(r,m) \in \intension{\kappa_j}$.    
Since $\kappa_1, \kappa_2, \ldots$ are mutually exclusive and   
exhaustive, there is exactly one action   
$\sfa_j$ with this    
property.   
\commentout{   
Joint protocols $\vec{P}$ and $\vec{P}'$ are {\em equivalent in context   
$\gamma$}, denoted $\vec{P} \approx_\gamma \vec{P}'$, if   
$P_i(\ell) = P'_i(\ell)$ for every local state $\ell=r_i(m)$ with   
$r \in \Rrep(\vec{P},\gamma)$.  That is, $\vec{P}$ and $\vec{P}'$ perform the   
same joint action in every global state that arises in   
$\Rrep(\vec{P}, \gamma)$.     
We remark that if $\vec{P} \approx_\gamma \vec{P}'$, then it easily   
follows that $\Rrep(\vec{P},\gamma) = \Rrep(\vec{P}',\gamma)$: we simply   
show by induction on $m$ that every prefix of a run in   
$\Rrep(\vec{P},\gamma)$ is a    
prefix of a run in $\Rrep(\vec{P}',\gamma)$, and vice versa.   
Finally, we say that}   
We are mainly interested in protocols that implement a kb program.   
Intuitively, a joint protocol $\vec{P}$ implements a kb program   
$\vec{\Pgkb}$ in context $\gamma$ if   
$\vec{P}$ performs the same actions as $\vec{\Pgkb}$   
in all runs of $\vec{P}$ that have positive probability,   
assuming that   
the knowledge tests in $\vec{\Pgkb}$ are interpreted with respect to    
the complete system $\PS$ extending $\Rrep(\vec{P},\gamma)$.     
Formally,    
a joint protocol $\vec{P}$ \emph{(de facto) implements} a   
joint kb   
program $\vec{\Pgkb}$ \cite{HM98a} in a context $\gamma = (\Gz, \vecmu)$ if    
$P_i(\ell) =    
\Pgkb^{\PS}_i(\ell)$ for every local state $\ell=r_i(m)$    
such that $r \in \Rrep(\vec{P},\gamma)$ and $\mu_i(r) \ne 0$,   
where $\PS$ is the complete system   
extending $\Rrep(\vec{P},\gamma)$.     
We remark   
that, in general, there may not be any joint protocols that    
implement a kb program in a given context, there may be exactly    
one, or there may be more than one (see \cite{FHMV} for examples).  This   
is somewhat analogous to    
the fact that there may not be any equilibrium of a game for some notions    
of equilibrium, there may be one, or there may be more than one.

\section{The Main Results}\label{sec:mainresults}   
   
Fix a game $\Gamma$ in normal form.   
Let $\Pnf_i$ be the protocol that, in   
initial state $s_S \in \Sigma^\Gamma_i$, chooses strategy $S$; let   
$\vPnf = (\Pnf_1, \ldots, \Pnf_n)$.     
\commentout{   
To explain the results, we first need to recall the notion of   
\emph{strong independence} \cite{BBD1,BBD2}.  Given a    
tuple $\vec{r} =   
(r^0, \ldots, r^{k-1})$ of numbers in    
$(0,1)^k$   
and an LPS   
$\vecmu = (\mu^0, \ldots, \mu^k)$, let $\vecmu \, \Box \, \vec{r}$   
be the probability measure    
$$(1 - r^0) \mu^0 + r^0[(1-r^1) \mu^1 + r^1[(1-r^2)\mu^2 + r^2[\cdots +   
r^{k-2}[(1-r^{k-1})\mu^{k-1} + r^{k-1}\mu^k)]\ldots ]]].$$   
The random variables $X_1, \ldots, X_h$ are strongly independent with   
respect to LPS $\vecmu$ if there exists a sequence    
$\vec{r}^j$, $j = 1, 2, \ldots$ of    
tuples in $(0,1)^k$     
such that   
$\vec{r}^j \rightarrow (0,\ldots, 0)$    
as $j\rightarrow\infty$,    
and $X_1, \ldots, X_h$ are   
independent with respect to $\vecmu \, \Box \, \vec{r}^j$ for $j = 1, 2, 3,   
\ldots$.}   
Let $\STRAT_i$ be the random   
variable on initial global states that associates with an initial global   
state~$s$ player $i$'s  strategy in $r$.     
As we said, Nash equilibrium arises in contexts with a common prior.   
Suppose that $\gamma = (\Gz,\mu)$ is a context with a common prior.    
We say that $\mu$ is \emph{compatible with}    
the mixed joint    
strategy $\vec{S}$   
if $\mu$ is the probability on pure joint strategies induced by $\vec{S}$   
(under the obvious identification of initial global states with joint   
strategies).    
   
\thm\label{thm:Nash1} The joint strategy $\vec{S}$ is a Nash equilibrium   
of the game    
$\Gamma$ iff there is a    
common prior probability measure $\mu$ on $\Gz^\Gamma$ such that $\STRAT_1,   
\ldots, \STRAT_n$ are independent with respect to $\mu$,   
$\mu$ is compatible with $\vec{S}$,   
and $\vPnf$ implements $\EQNF^\Gamma$ in the context   
$(\Gz^\Gamma,\mu)$.    
\ethm   
   
\prf    
Suppose that    
$\vec{S}$ is a (possibly mixed strategy) Nash equilibrium   
of the game $\Gamma$.     
Let $\mu_{\vec{S}}$ be the unique probability on $\Gz^\Gamma$ compatible   
with $\vec{S}$.      
If $\vec{S}$ is played, then the probability of a run where the    
pure joint strategy $(T_1, \ldots, T_n)$ is played    
is just the product of the probabilities assigned   
to $T_i$ by $S_i$, so $\STRAT_1, \ldots, \STRAT_n$ are independent with   
respect to $\mu_{\vec{S}}$.     
To see that $\vPnf$ implements $\EQNF^\Gamma$ in the context   
$\gamma = (\Gz^\Gamma,\mu_{\vec{S}})$,    
let $\ell = r_i(0)$ be a local state such that $r = \Rrep(\vPnf,\gamma)$   
and $\mu(r) \ne 0$.  If $\ell = s_T$, 
then $\Pnf_i(\ell) = T$, so
$T$ must be in the support of $S_i$.  Thus,    
$T$ must be a best response to $\vec{S}_{-i}$, the joint strategy   
where each player $j \ne i$ plays its component of   
$\vec{S}$.   
Since $i$ uses strategy $T$ in $r$, the formula    
$B_i(\doact_i(T'))$ holds at $(r,0)$ iff $T' = T$.  Moreover,    
since $T$ is a best response, if $u$ is $i$'s expected utility with the   
joint strategy $\vec{S}$, then for all $T'$, the formula   
$\doact_i(T') \RCond (\EU_i \le u)$ holds at $(r,0)$.  Thus,    
$(\EQNF^\Gamma_i)^\PS(\ell) = T$, where $\PS$ is the complete system   
extending $\Rrep(\vPnf,\gamma)$.  It follows that $\vPnf$ implements   
$\EQNF^\Gamma$.

For the converse, suppose that $\mu$ is a common prior probability   
measure on $\Gz^\Gamma$, $\STRAT_1, \ldots, \STRAT_n$ are independent   
with respect to $\mu$, $\mu$ is compatible with   
$\vec{S}$, and $\vPnf$ implements $\EQNF^\Gamma$ in the context   
$\gamma = (\Gz^\Gamma,\mu)$.    
We want to show that $\vec{S}$ is a Nash equilibrium.  It suffices to   
show that each pure strategy $T$ in the support of $S_i$ is a best   
response to $\vec{S}_{-i}$.   Since $\mu$ is compatible with $\vec{S}$,   
there must be a run $r$ such that $\mu(r) > 0$ and $r_i(0) = s_T$   
(i.e., player $i$ chooses $T$ in run $r$).  It since $\vPnf$   
implements $\EQNF^\Gamma$, and  in the context $\gamma$,    
$\EQNF^\Gamma$ ensures that no deviation from $T$ can improve $i$'s   
expected utility with respect to $\vec{S}_{-i}$,    
it follows that $T$ is indeed a best response.      
\eprf   

\commentout{   

\thm\label{thm:perfeq} The joint strategy $\vec{S}$ is a perfect   
equilibrium of the game    
$\Gamma$ iff there is a common prior LPS $\vecnu$ on $\Gz^\Gamma$ such that   
$\STRAT_1,   
\ldots, \STRAT_n$ are strongly independent with respect to $\vecnu$,   
$\vPnf$ is compatible with $\vec{S}$, and $\vPnf$ implements    
$\EQNF^\Gamma$ in context $(\Gz^\Gamma,\vecnu)$.    
\ethm   
   
\prf STILL TO COME \eprf   
   
[[SHOULD WE TALK ABOUT PROPER EQUILIBRIUM?  PERHAPS EASIER WOULD BE TO   
COMPARE OUR RESULTS TO THOSE OF BBD HERE, AND SAY THAT WE CAN SIMILARLY   
CHARACTERIZE PROPER EQUILIBRIUM.  THEN SAY THAT BBD DIDN'T CONSIDER   
RATIONALIZABILITY.]]   
   
[Yoram - I support your suggestion here. There is little added    
value in redoing for proper without any novel aspect. We could state    
a proposition without proof that would claim the same result for proper,    
without formal details, or just state it in the running discussion text.]   
}   
As is well known, players can sometimes achieve better outcomes than a   
Nash equilibrium    
if they have access to a helpful mediator.    
Consider the   
simple 2-player game described in Figure~\ref{table1}, where Alice,    
the row player, must choose between top and bottom ($T$ and $B$), while   
Bob, the column player, must choose between left and right ($L$ and $R$):   
   
\begin{figure}[htb]   
\begin{center}   
\begin{tabular}{l|c|c|}\multicolumn{1}{l}{}&\multicolumn{1}{c}{$L$}&\multicolumn{1}{c}{$R$}\\     
\cline{2-3}    
 $T$& $(3,3)$& $(1,4)$ \\ \cline{2-3}   
 $B$&$(4,1)$&$(0,0)$\\ \cline{2-3}   
\multicolumn{1}{l}{\vspace{-3mm}}   
\end{tabular}\\   
\caption{A simple 2-player game.}\label{table1}   
\end{center}   
\end{figure}   
It is not hard to check that the best Nash equilibrium for this game has   
Alice randomizing between $T$ and $B$, and Bob randomizing   
between $L$ and $R$; this gives    
each   
of them expected utility 2.  They   
can do better with a trusted mediator, who makes a   
recommendation by choosing at random between $(T,L)$, $(T,R)$, and   
$(B,L)$.     
This gives each of them expected utility $8/3$.  This is a   
\emph{correlated equilibrium} since, for example, if the mediator   
chooses $(T,L)$, and thus sends recommendation $T$ to Alice and $L$ to   
Bob,    
then   
Alice considers it equally likely that Bob was told $L$ and $R$,   
and thus has no incentive to deviate; similarly, Bob has no incentive to   
deviate.  In general, a distribution $\mu$ over pure joint strategies is a   
correlated equilibrium if players cannot do better than following a   
mediator's recommendation if a mediator makes recommendations according   
to $\mu$.  (Note that, as in our example, if a mediator chooses a joint   
strategy $(S_1, \ldots, S_n)$ according to $\mu$, the mediator   
recommends $S_i$ to player $i$; player $i$ is not told the joint   
strategy.)  We omit the formal definition of correlated equilibrium (due   
to Aumman \citeyear{Aumann74}) here; however, we stress that a   
correlated equilibrium is a distribution over (pure) joint strategies.   
We can easily capture correlated equilibrium using $\EQNF$.   
   
\thm\label{thm:correlated} The distribution $\mu$ on joint strategies is   
a correlated equilibrium of the game    
$\Gamma$ iff $\vPnf$ implements $\EQNF^\Gamma$ in the context   
$(\Gz^\Gamma,\mu)$.    
\ethm

Both Nash equilibrium and correlated equilibrium require a common prior   
on runs.     
By dropping this assumption, we    
get another standard solution concept: \emph{rationalizability}   
\cite{Ber84,Pearce84}.  Intuitively, a strategy for player $i$ is   
rationalizable if it is a best response to some beliefs that player $i$    
may have about    
the strategies that other players are following,   
assuming that these strategies are themselves best responses to beliefs   
that the other players have about strategies that other players are   
following, and so on.   
To make this precise, we need a little notation.     
Let $\S_{-i} = \Pi_{j \ne i} \S_j$.   Let $u_i(\vec{S})$ denote player   
$i$'s utility if the strategy tuple $\vec{S}$ is played.   
We describe player $i$'s beliefs about what strategies the other players   
are using by a probability $\mu_i$ on $\S_{-i}$.     
A strategy $S$ for player $i$ is a \emph{best response   
to beliefs described by a probability $\mu_i$ on    
$\S_{-i}(\Gamma)$} 
if $\sum_{\vec{T} \in \S_{-i}} u_i(S,\vec{T})   
\mu_i(\vec{T}) \ge  \sum_{\vec{T} \in \S_{-i}} u_i(S',\vec{T})   
\mu_i(\vec{T})$ for all $S' \in \S_i$.     
Following Osborne and Rubinstein \citeyear{OR94}, we say that a strategy $S$   
for player $i$ in game $\Gamma$  
is \emph{rationalizable} if, for each player $j$, there is a set $Z_j   
\subseteq \S_j(\Gamma)$ and, for each strategy $T \in Z_j$,  a   
probability measure $\mu_{j,T}$ on $\S_{-j}(\Gamma)$ whose support is   
$Z_{-j}$ such that     
\begin{itemize}   
\item $S \in Z_i$; and   
\item for each player $j$ and strategy $T \in Z_j$, $T$ is a best   
response to the beliefs $\mu_{j,T}$.   
\end{itemize}   

For ease of exposition, we consider only    
pure    
rationalizable strategies.  This is essentially without loss of generality.     
It is easy to see that a mixed strategy $S$ for player $i$ is a best   
response to some beliefs $\mu_i$ of player $i$ iff each pure   
strategy in the support of $S$ is a best response to $\mu_i$.   
Moreover, we can assume without loss of generality that   
the support of $\mu_i$ consists of only pure joint strategies.   
   
\thm\label{thm:rationalizability}    
A pure strategy $S$ for player $i$ is 
rationalizable iff there    
exist probability measures $\mu_1, \ldots, \mu_n$, a set $\Gz   
\subseteq \Gz^\Gamma$, and 
a state 
$\vec{s} \in \Gz$   
such that $\Pnf_i(s_i) = S$ and $\vPnf$ implements   
$\EQNF^\Gamma$ in the context $(\Gz,\vecmu)$.    
\ethm   
   
\prf    
First, suppose that $\vPnf$ implements $\EQNF^\Gamma$ in context
$(\Gz,\vecmu)$.  We show that for each state $\vec{s} \in \Gz$ and
player $i$, the strategy $S_{\vec{s},i} = \vPnf_i(s_i)$ is
rationalizable.  
Let $Z_i = \{S_{\vec{s},i}: \vec{s} \in \Gz\}$. For 
$S \in Z_i$, let $E(S) = \{\vec{s}  \in \Gz: s_i = s_S\}$; that
is, $E(S)$ consists  consists of all initial global states where player
$i$'s local state is $s_S$; let $\mu_{i,S} =    
\mu_i( \cdot \mid E(S))$ (under the obvious identification of global states   
in $\Gz$ with joint strategies).   
Since $\vPnf$ implements $\EQNF^\Gamma$, it easily follows that $S$ 
best response to $\mu_{i,S}$.  Hence, all the strategies in $Z_i$   
are rationalizable, as desired.

For the converse, let $Z_i$ consist of all the pure rationalizable strategies 
for player $i$.  It follows from the definition of rationalizability   
that, for each strategy $S \in Z_i$, there exists a probability measure   
$\mu_{i,S}$ on $Z_{-i}$ such that $S$ is a best response to $\mu_{i,S}$.   
For a set~$Z$ of strategies, we denote by $\tilde{Z}$ the set 
$\{s_T: T\in Z\}$. 
Set $\Gz = \tilde{Z_1} \times \ldots \times \tilde{Z_n}$, 
and choose some measure   
$\mu_i$ on $\Gz$ such that $\mu_i( \cdot \mid E(S)) = \mu_{i,S}$ for    
all $S \in Z_i$.  (We can take $\mu_i = \sum_{S \in Z_i} \alpha_S   
\mu_{i,S}$, where $\alpha_S \in (0,1)$ and    
$\sum_{S \in Z_i} \alpha_S =1$.)     
Recall that $\Pnf_i(s_S)=S$ for all states $s_S$. 
It immediately follows that, for every rationalizable joint strategy 
$\vec{S}= (S_1,\ldots,S_n)$, both 
$\vec{s}=(s_{S_1},\ldots,s_{S_n})\in\Gz$, and 
$\vec{S}=\vPnf(\vec{s})$. 
Since the states in~$\Gz$ all correspond to rationalizable 
strategies, and by definition of rationalizability each (individual)
strategy~$S_i$ is a best response to $\mu_{i,S}$, 
it is easy to check that $\vPnf$ implements $\EQNF^\Gamma$ in   
the context $(\Gz^\Gamma,\vecmu)$, as desired. \eprf  

We remark that    
Osborne and Rubinstein's definition of rationalizability allows   
$\mu_{j,T}$ to be such that $j$ believes that other players' strategy   
choices are correlated.  In most of the literature, players are assumed   
to believe that other players' choices are made independently.     
If we add that requirement, then we must impose the same requirement on   
the probability measures $\mu_1, \ldots, \mu_n$ in   
Theorem~\ref{thm:rationalizability}.    

\begin{figure}[ht]   
\centerline{\psfig{figure=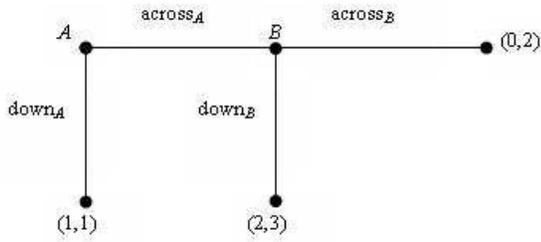,height=1.4in}}   
\caption{A game with an unreasonable Nash equilibrium.}   
\label{fig:game1}   
\end{figure}

Up to now we have considered solution concepts for games in normal form.   
Perhaps the best-known solution concept for games in extensive form is   
\emph{sequential equilibrium} \cite{KW82}.     
Roughly speaking, a joint strategy $\vec{S}$ is a sequential equilibrium   
if $S_i$ is a best response to $\vec{S}_{-i}$ at \emph{all} information   
sets, not just the information sets that are reached with positive   
probability when playing $\vec{S}$.  To understand how sequential   
equilibrium differs from Nash equilibrium,    
consider the game shown in Figure~\ref{fig:game1}.   
   
One Nash equilibrium of this game has $A$ playing down$_A$ and $B$   
playing across$_B$.     
However, this is not a sequential equilibrium,   
since playing across is not a best response for $B$ if $B$ is called on   
to play.     
This is not a problem in a Nash equilibrium   
because the node where $B$ plays is not reached in the equilibrium.   
Sequential equilibrium refines Nash equilibrium (in the sense that every   
sequential equilibrium is a Nash equilibrium) and does not allow   
solutions such as (down$_A$, across$_B$). Intuitively, in a   
sequential equilibrium, every player must make a best response at   
every information set (even if it is reached with probability 0). In   
the game shown in Figure~\ref{fig:game1}, the unique joint strategy   
in a sequential equilibrium has $A$ choosing   
across$_A$ and $B$ choosing down$_B$.     

The main difficulty in defining sequential equilibrium lies in capturing
the intuition of best response in information sets that are reached with
probability 0.  To deal with this, a sequential equilibrium is defined
to be a pair $(\vec{S},\beta)$, consisting of a joint strategy
$\vec{S}$ and a \emph{belief system} $\beta$, which associates with every 
information set $I$ a probability $\beta(I)$ on the histories in $I$.
There are a number of somewhat subtle consistency conditions on these pairs
pairs; we omit them here due to lack of space (see
\cite{KW82,OR94} for details).   Our result depends on a
recent characterization of sequential equilibrium \cite{Hal36} that uses
nonstandard probabilities, which can assign infinitesimal probabilities
to histories.  By assuming that every history gets positive (although
possibly infinitesimal) 
probability, 
we can avoid the problem of
dealing with information sets that are reached with probaility 0.
   
\ijcai{It is well known that to}
\ijcaionly{To} every nonstandard   
real number $r$, there is a closest standard real number denoted   
$\stand{r}$,
and read ``the standard part of~$r$'':
$|r-\stand{r}|$ is an infinitesimal.     
Given a   
nonstandard probability measure $\nu$, we can define the standard   
probability measure $\stand{\nu}$ by taking $\stand{\nu}(w) =   
\stand{\nu(w)}$.     
A nonstandard probability $\nu$ on $\Gz$ is \emph{compatible with} joint   
strategy $\vec{S}$ if $\stand{\nu}$ is the probability on pure    
strategies induced by $\vec{S}$.   When dealing with nonstandard   
probabilities, we generalize the definition of implementation by   
requiring only that $\vec{P}$ performs the same actions as $\vec{\Pgkb}$    
in runs~$r$ of $\vec{P}$ such that $\stand{\nu}(r) > 0$.     
Moreover, the expression ``$EU_i = x$'' in $\EQEF^\Gamma$ is interpreted as
``the standard part of $i$'s expected utility is $x$'' (since $x$ ranges over
the standard real numbers).
   
\thm\label{thm:seqeq}    
If $\Gamma$ is a game with perfect recall%
\footnote{These are games where players remember all actions made and   
the states they have gone through; we give a formal definition in the   
full paper.  See also \cite{OR94}.}   
there is a belief system $\beta$ such that $(\vec{S},\beta)$ 
is a sequential   
equilibrium of $\Gamma$    
iff there is a common prior nonstandard   
probability    
measure   
$\nu$ on $\Gz^\Gamma$ that gives positive measure to all   
states such that $\STRAT_1, \ldots, \STRAT_n$ are     
independent with respect    
to $\nu$, $\nu$ is compatible with $\vec{S}$, and $\vPef$   
implements $\EQEF^\Gamma$ in the standard context $(\Gz^\Gamma,\nu)$.    
\ethm   
   
This is very similar in spirit to Theorem~\ref{thm:Nash1}.  The key   
difference is the use of a nonstandard probability measure.   
Intuitively, this forces $\vec{S}$ to be a best response even at   
information sets that are reached with (standard) probability 0.   

The effect of interpreting ``$\EU_i = x$'' as ``the standard part of 
$i$'s expected utility is $x$'' is that we ignore infinitesimal
differences.  Thus, for example, the strategy $\vPef_i(\vec{s_0})$ might
not be a best response to $\vec{S}_{-i}$; it might just be an
$\epsilon$-best response for some infinitesimal $\epsilon$.  
As we show in the full paper, it follows from Halpern's
\citeyear{Hal36} results that
we can also obtain a characterization of \emph{(trembling hand) perfect 
equilibrium} \cite{Selten75}, another standard refinement of Nash equilibrium, 
if we interpret ``$\EU_i = x$'' as ``the expected
utility for agent $i$ is $x$'' and allow $x$ to range over the
nonstandard reals instead of just the standard reals. 
\section{Conclusions}\label{sec:conclusions}   
We have shown how a number of different solution concepts    
from game theory can be captured by essentially one knowledge-based   
program, which comes in two variants: one appropriate for normal-form   
games and one for extensive-form games.  The differences between these   
solution concepts is captured by changes in   
the context in which the games are played: whether players have a 
common prior   
(for Nash equilibrium, correlated equilibrium, and sequential   
equilibrium) or not (for rationalizability), whether strategies are   
chosen independently (for Nash equilibrium, sequential equilibrium, and   
rationalizability) or not 
\ijcai{
(for correlated equilibrium; we can also allow this for rationalizability),}
\ijcaionly{(for correlated equilibrium);}
and whether uncertainty is represented   
using a 
\ijcai{
standard probability measure (for Nash equilibrium, correlated   
equilibrium, and rationalizability) or  a    
nonstandard probability measure (for sequential equilibrium).     
}
\ijcaionly{standard or nonstandard probability measure.}

Our results can be viewed as showing that each of these solution   
concepts \emph{sc} can be characterized in terms of common knowledge of   
rationality (since the kb programs $\EQNF^\Gamma$ and   
$\EQEF^\Gamma$ embody rationality, and we are interested in systems   
``generated'' by these program, so that rationality holds at all   
states), and common knowledge of some other features $X_{sc}$   
captured by the context appropriate for \emph{sc} (e.g., that strategies   
are chosen independently or that the prior).  Roughly speaking, our   
results say that    
if $X_{sc}$ is common knowledge in a system,    
then common knowledge of rationality implies that the   
strategies used must satisfy solution concept \emph{sc}; conversely, if   
a joint strategy $\vec{S}$    
satisfies \emph{sc}, then there is a system where $X_{sc}$ is   
common knowledge, rationality is common knowledge, and $\vec{S}$ is   
being played at some state.    
Results similar in spirit have been proved for    
rationalizability \cite{BD87a} and correlated equilibrium   
\cite{Aumann87}.     
\commentout{   
There are also epistemic characterizations of   
rationalizability in terms of common knowledge of rationality   
\cite{TW88} and a characterization of necessary and sufficient   
conditions on knowledge of rationality needed for  Nash equilibrium   
\cite{AB95}.   
}   
Our approach allows us to unify and extend these results and,
as    
suggested in the introduction, applies even to settings
where the game is not common knowledge and
in settings    
where uncertainty is not represented by probability.     
We believe that the approach captures the essence of the intuition that    
a solution concept should embody common knowledge of rationality.   

\bibliography{z,joe}   
\bibliographystyle{named}   
\end{document}